\documentclass[a4paper]{jpconf}
\usepackage{graphicx}
\usepackage{epsf}

\usepackage[]{latexsym}
\usepackage{bm}

\newcommand{\be}{\begin{equation}}\newcommand{\ee}{\end{equation}}
\newcommand{\bea}{\begin{eqnarray}}\newcommand{\eea}{\end{eqnarray}}
\newcommand{\brr}{\begin{array}}\newcommand{\err}{\end{array}}
\newcommand{\bit}{\begin{itemize}}\newcommand{\eit}{\end{itemize}}
\newcommand{\ben}{\begin{enumerate}}\newcommand{\een}{\end{enumerate}}

\newcommand{\ba}{\begin{array}}
\newcommand{\ea}{\end{array}}

\def\lan{\langle}
\def\lf{\left}

\def\non{\nonumber}\def\ran{\rangle}

\def\ri{\right}
\def\al{\alpha}\def\bt{\beta}

\def\te{\theta}
\def\si{\sigma}
\def\om{\omega}

\def\1{{_{1}}}\def\2{{_{2}}}

\newcommand{\ide}{1\hspace{-1mm}{\rm I}}

\def\noHe0{:\;\!\!\;\!\!:H_e(0):\;\!\!\;\!\!:}
\def\noHm0{:\;\!\!\;\!\!:H_\mu(0):\;\!\!\;\!\!:}

\def\vect#1{{\bm #1}}

\def\lan{\langle}
\def\lf{\left}

\def\non{\nonumber}
\def\ran{\rangle}

\def\ri{\right}

\def\al{\alpha}\def\bt{\beta}

\def\te{\theta}

\def\si{\sigma}
\def\om{\omega}

\def\1{{_{1}}}\def\2{{_{2}}}

\begin{document}
\title{Physical  flavor neutrino states}

\author{Massimo Blasone$^{\dag \, 1,2}$}

\address{$^{1}$ Dipartimento di Matematica e Informatica,
Universit\`a degli Studi di Salerno, Via Ponte don Melillo,
I-84084 Fisciano (SA), Italy
\\
  $^{2}$ INFN Sezione di Napoli,
Gruppo collegato di Salerno, Baronissi (SA), Italy
}

\ead{$^{\dag}$blasone@sa.infn.it}

\begin{abstract}
 The problem of representation for flavor states of mixed neutrinos is discussed.
By resorting to recent results, it is shown that a specific representation exists in which a number of conceptual problems are resolved.
Phenomenological consequences of
our analysis are explored.
\end{abstract}

\section{Introduction}

The  detailed study of neutrino mixing in the context of quantum field theory has led
to  the discovery of unexpected features associated to such a phenomenon.
Indeed, it was found \cite{BV95} that the Hilbert space where the mixed (flavor) field
operators are defined is unitarily inequivalent, in the infinite volume
limit, to the Hilbert space for the original (unmixed) field
operators. This is due to the condensate structure of the vacuum state for the flavor fields, the flavor vacuum, which turns out to be a coherent state.

These results have been then found to have general validity independently of the type or number of field involved \cite{FHY99,Blasone:2001du,Ji,hannabuss,Blasone:2003hh}. Flavor oscillation formulas were derived \cite{BHV99,BJV01,3flav} exhibiting corrections with respect to the usual ones derived in quantum mechanics \cite{Pontecorvo}. Effects of flavor vacuum structure have been discussed also in connection with cosmological constant \cite{Blasone:2004yh} and Lorentz invariance \cite{BlaMagueijo}. Flavor states have also been studied as (relativistic) examples of single-particle entanglement \cite{Ent}. 
More recently, a non-abelian gauge structure has been recognized to be associated to flavor mixing \cite{Blasone:2010zn}.

In this paper, we discuss a novel representation of flavor vacuum and flavor states which presents some advantages with respect to the one previously adopted \cite{BV95}. Phenomenological consequences of this choice are explored.

\section{Neutrino mixing in QFT\label{standardmix}}

 Let us denote by
$\nu_{e}$, $\nu_{\mu}$ the neutrino fields with definite flavors and
by $\nu_{1}$, $\nu_{2}$ the neutrino fields with definite masses
$m_{1}$, $m_{2}$, respectively.
We consider the lagrangian
\begin{equation}\label{lag2}
{\cal L} = {\bar \nu}_e\left ( i\!\not\!\partial -m_{e}\right)\nu_e +
{\bar    \nu}_\mu\left(i\!\not\!\partial - m_{\mu}\right)\nu_\mu   - \;
m_{e \mu} \;\left({\bar \nu}_{e}   \nu_{\mu} + {\bar \nu}_{\mu}
\nu_{e}\right)
\;.\end{equation}
which is diagonalized by the
transformation
\begin{eqnarray}
\nu_{e}(x)   &=&\nu_{1}(x)\,\cos\theta +    \nu_{2}(x)\,\sin\theta\, ,
\\ [2mm]
\label{rot1}
\nu_{\mu}(x) &=&-\nu_{1}(x)\,\sin\theta    + \nu_{2}(x)\,\cos\theta
\,,\end{eqnarray}
where $\theta$ is the mixing angle and
$ m_{e} = m_{1}\cos^{2}\theta +
m_{2} \sin^{2}\theta~$,  $m_{\mu} = m_{1}\sin^{2}\theta + m_{2}
\cos^{2}\theta~$,   $m_{e\mu} =(m_{2}-m_{1})\sin\theta \cos\theta\,$.
The result is
the sum of two free Dirac Lagrangians:
\bea\label{Lagrmass} {\cal L} &=& {\bar \nu}_1 \lf( i
\not\!\partial -
  m_{1}\ri)\nu_1  +  {\bar \nu}_2 \lf( i \not\!\partial -
  m_{2}\ri)\nu_2\,.  \eea
The expansions for $\nu_1$ and $\nu_2$ are:
\begin{equation}\label{2.2}\nu_{i}(x) =
\sum_{r=1,2}
\int \frac{d^3  k}{(2\pi)^\frac{3}{2}} \left[u^{r}_{{\bf k},i}
\alpha^{r}_{{\bf k},i}(t)\:+    v^{r}_{-{\bf k},i}\beta^{r\dag }_{-{\bf
k},i}(t)   \right] e^{i {\bf k}\cdot{\bf x}} , \qquad i=1,2 \,,
\end{equation}
with $\alpha^{r}_{{\bf k},i}(t)=e^{-i\omega_{i} t}\alpha^{r}_{{\bf
k},i}(0)$, $\beta^{r}_{{\bf k},i}(t)=e^{-i\omega_{i} t}\beta^{r}_{{\bf
k},i}(0)$ and  $\omega_{i}=\sqrt{{\bf k}^2+m_i^2}$.  Here and in the
following we use $t\equiv x_0$, when no misunderstanding  arises.   The
vacuum for the $\alpha_i$ and $\beta_i$ operators is denoted by
$|0\rangle_{1,2}$:  $\; \; \alpha^{r}_{{\bf k},i}|0\rangle_{12}=
\beta^{r }_{{\bf    k},i}|0\rangle_{12}=0$.   The anticommutation
relations are the usual ones (see Ref.~\cite{BV95}).  The orthonormality
and  completeness relations are:
\begin{equation}
u^{r\dag}_{{\bf k},i} u^{s}_{{\bf k},i} =   v^{r\dag}_{{\bf k},i}
v^{s}_{{\bf k},i} = \delta_{rs}  \;,\quad  u^{r\dag}_{{\bf k},i}
v^{s}_{-{\bf k},i} =  v^{r\dag}_{-{\bf k},i} u^{s}_{{\bf k},i} =
0\;,\quad \sum_{r}(u^{r}_{{\bf k},i} u^{r\dag}_{{\bf k},i} +
v^{r}_{-{\bf k},i} v^{r\dag}_{-{\bf k},i}) = \ide \;.
\end{equation}

The fields $\nu_e$ and $\nu_\mu$ are thus    completely determined
through    Eq.~(\ref{rot1}), which can be rewritten in the following
form (we use $(\sigma,j)=(e,1) , (\mu,2)$):
\begin{eqnarray}\label{exnue1}
\nu_{\sigma}(x)     &=& G^{-1}_{\theta}(t)\, \nu_{j}(x)\, G_{\theta}(t)
= \sum_{r=1,2}
\int \frac{d^3  k}{(2\pi)^\frac{3}{2}}   \left[ u^{r}_{{\bf k},j}
\alpha^{r}_{{\bf k},\sigma}(t) +    v^{r}_{-{\bf k},j}
\beta^{r\dag}_{-{\bf k},\sigma}(t) \right]  e^{i {\bf k}\cdot{\bf x}},
\\ \label{BVgen}
G_{\theta}(t) &=& \exp\left[\theta \int d^{3}{\bf x}
\left(\nu_{1}^{\dag}(x)\nu_{2}(x) -    \nu_{2}^{\dag}(x)
\nu_{1}(x)\right)\right]\, ,
\end{eqnarray}
where $G_{\theta}(t)$    is the generator of the mixing transformations
(\ref{rot1}) (see  Ref.~\cite{BV95} for a discussion of its properties).

Eq.(\ref{exnue1}) gives an expansion of the flavor fields $\nu_{e}$ and
$\nu_{\mu}$ in the same basis of $\nu_{1}$ and $\nu_{2}$.  The flavor annihilation operators are then identified with
\begin{equation}\label{BVoper}
\left(\begin{array}{c} \alpha^{r}_{{\bf k},\sigma}(t)\\
\beta^{r\dag}_{{-\bf k},\sigma}(t)
\end{array}\right)
= G^{-1}_{\theta}(t)  \left(\begin{array}{c} \alpha^{r}_{{\bf k},j}(t)\\
\beta^{r\dag}_{{-\bf k},j}(t)
\end{array}\right)
G_{\theta}(t)\,.
\end{equation}
The action of the mixing generator on the vacuum $|0 \ran_{1,2}$ is
non-trivial and we have:
\be\label{2.22} |0 (t)\ran_{e,\mu}\equiv G^{-1}_{\te}(t)\; |0
\ran_{1,2}\;. \ee
$|0(t) \ran_{e,\mu}$ is the {\em flavor vacuum}, i.e.,  the vacuum for
the flavor fields\cite{BV95}.

The explicit expression of the flavor annihilation operators is (we
choose ${\bf k}=(0,0,|{\bf k}|)$):
\begin{equation}\label{BVmatrix}
\left(\begin{array}{c} {\alpha}^{r}_{{\bf k},e}(t)\\ {\alpha}^{r}_{{\bf
k},\mu}(t)\\ {\beta}^{r\dag}_{{-\bf k},e}(t)\\ {\beta}^{r\dag}_{{-\bf
k},\mu}(t)
\end{array}\right)
\, = \, \left(\begin{array}{cccc} c_\theta &  s_\theta\, |U_{{\bf k}}|
&0 & s_\theta \, \epsilon^{r} \,|V_{{\bf k}}| \\ - s_\theta\, |U_{{\bf
k}}|& c_\theta
&  s_\theta \,\epsilon^{r} \,|V_{{\bf k}}| & 0  \\ 0& - s_\theta
\,\epsilon^{r} \,|V_{{\bf k}}|
&c_\theta & s_\theta \,|U_{{\bf k}}| \\ - s_\theta \,\epsilon^{r}
\,|V_{{\bf k}}| & 0 &
- s_\theta\, |U_{{\bf k}}| & c_\theta \\
\end{array}\right)
\left(\begin{array}{c} \alpha^{r}_{{\bf k},1}(t)\\ \alpha^{r}_{{\bf
k},2}(t)\\ \beta^{r\dag}_{{-\bf k},1}(t)\\ \beta^{r\dag}_{{-\bf k},2}(t)
\end{array}\right)\,,
\end{equation}
where $c_\theta\equiv \cos\theta$, $s_\theta\equiv \sin\theta$,
$\epsilon^r\equiv (-1)^r$, and
\begin{eqnarray}
&&|U_{{\bf k}}|\equiv u^{r\dag}_{{\bf k},2}u^{r}_{{\bf k},1}=
v^{r\dag}_{-{\bf k},1}v^{r}_{-{\bf k},2}\,, \\
\label{2.37}
&&|V_{{\bf k}}|\equiv \epsilon^{r}\;u^{r\dag}_{{\bf  k},1}v^{r}_{-{\bf
k},2}= -\epsilon^{r}\;u^{r\dag}_{{\bf k},2}v^{r}_{-{\bf k},1}\,.
\end{eqnarray}
The number of particles with definite mass condensed in
the flavor vacuum is given by
\bea\label{2.41b} &&\;_{e,\mu}\lan 0(t)| \al_{{\bf k},i}^{r \dag}
\al^r_{{\bf k},i} |0(t)\ran_{e,\mu}=  \;_{e,\mu}\lan 0(t)| \bt_{{\bf k},i}^{r \dag}
\bt^r_{{\bf k},i} |0(t)\ran_{e,\mu}= \sin^{2}\te\; |V_{{\bf
k}}|^{2} , \qquad i=1,2\,. \eea

Following the usual procedure, we define the flavor charges for the flavor neutrino fields \cite{BJV01}:
\bea \label{flavcharges} && Q_{\sigma}(t) \,= \,\int d^{3}{\bf
x}\,
 \nu_{\sigma}^{\dag}(x)\;\nu_{\sigma}(x)~,
\qquad \sigma=e,\mu,
\eea
with $Q_{e}(t) + Q_{\mu}(t) = Q$, where $Q$ is the total (conserved) lepton charge.

The flavor charges are diagonal when expressed in terms
of the flavor ladder operators Eq~(\ref{BVmatrix}):
\bea
{Q}_{\sigma}(t)=\sum_r \int
d^3\mathbf{k}\lf({\alpha}^{r \dagger}_{\mathbf{k}\sigma}(t)
{\alpha}^r_{\mathbf{k}\sigma}
(t)- {\beta}^{r \dagger}_{-\mathbf{k}\sigma}(t)
{\beta}^r_{-\mathbf{k}\sigma}(t)\ri).\eea
The eigenstates of such charge operators (flavor neutrino states)  are consistently defined as:
\bea|{\nu}^{\, r}_{\mathbf{k}\sigma}(t)\ran \equiv
{\alpha}^{r \dagger}_{\mathbf{k}\sigma}(t)|0(t)\ran_{e,\mu}.\eea

\section{General expansion}

It was  noted \cite{FHY99},  that   in the expansion Eq.~(\ref{exnue1}) one could  use
eigenfunctions with arbitrary masses $\mu_\sigma$, and therefore not
necessarily the same as the masses which appear in the (diagonalized) Lagrangian.  On
this basis, the flavor fields can be also written as \cite{FHY99,remarks}
\begin{eqnarray}\label{exnuf2}
\nu_{\sigma}(x)     &=& \sum_{r=1,2}
\int \frac{d^3  k}{(2\pi)^\frac{3}{2}}  \left[
u^{r}_{{\bf k},\sigma} {\widetilde \alpha}^{r}_{{\bf k},\sigma}(t) +
v^{r}_{-{\bf k},\sigma} {\widetilde \beta}^{r\dag}_{-{\bf k},\sigma}(t)
\right]  e^{i {\bf k}\cdot{\bf x}} ,
\end{eqnarray}
where $u_{\sigma}$ and $v_{\sigma}$ are the helicity eigenfunctions with
mass $\mu_\sigma$. We denote by a tilde the generalized flavor operators
introduced in Ref.~\cite{FHY99} in order to distinguish them from the
ones defined in Eq.~(\ref{BVoper}).  The expansion Eq.~(\ref{exnuf2}) is
more general than the one in Eq.~(\ref{exnue1}) since the latter
corresponds to the particular choice $\mu_e\equiv m_1$, $\mu_\mu \equiv
m_2$.

The relation between the flavor and the mass operators is now:
\begin{equation}\label{FHYoper}
\left(\begin{array}{c} {\widetilde \alpha}^{r}_{{\bf k},\sigma}(t)\\
{\widetilde \beta}^{r\dag}_{{-\bf k},\sigma}(t)
\end{array}\right)
= K^{-1}_{\theta,\mu}(t)  \left(\begin{array}{c} \alpha^{r}_{{\bf
k},j}(t)\\ \beta^{r\dag}_{{-\bf k},j}(t)
\end{array}\right)
K_{\theta,\mu}(t) ~~,
\end{equation}
with $(\sigma,j)=(e,1) , (\mu,2)$ and
where $K_{\theta,\mu}(t)$ is the generator of the transformation
(\ref{rot1}) and can be written as
\begin{eqnarray}\label{FHYgen}
K_{\theta,\mu}(t)&=& I_{\mu}(t)\, G_{\theta}(t) \, ,
\\ \label{Igen}
I_{\mu}(t)&=& \prod_{{\bf k}, r}\, \exp\left\{ i
\mathop{\sum_{(\sigma,j)}} \xi_{\sigma,j}^{\bf k}\left[
\alpha^{r\dag}_{{\bf k},j}(t)\beta^{r\dag}_{{-\bf k},j}(t) +
\beta^{r}_{{-\bf k},j}(t)\alpha^{r}_{{\bf k},j}(t) \right]\right\}\,,
\end{eqnarray}
with $\xi_{\sigma,j}^{\bf k}\equiv (\chi_\sigma - \chi_j)/2$ and
$\cot\chi_\sigma = |{\bf k}|/\mu_\sigma$, $\cot\chi_j = |{\bf k}|/m_j$.
For $\mu_e\equiv m_1$, $\mu_\mu \equiv m_2$ one has $I_{\mu}(t)=1$.
The explicit  matrix form of the flavor operators is \cite{FHY99}:
\begin{equation}\label{FHYmatrix}
\left(\begin{array}{c} {\widetilde \alpha}^{r}_{{\bf k},e}(t)\\
{\widetilde \alpha}^{r}_{{\bf k},\mu}(t)\\ {\widetilde
\beta}^{r\dag}_{{-\bf k},e}(t)\\ {\widetilde \beta}^{r\dag}_{{-\bf
k},\mu}(t)
\end{array}\right)
\, = \, \left(\begin{array}{cccc} c_\theta\, \rho^{\bf k}_{e1} &
s_\theta \,\rho^{\bf k}_{e2}  &i c_\theta \,\lambda^{\bf k}_{e1} & i
s_\theta \,\lambda^{\bf k}_{e2}  \\ - s_\theta \,\rho^{\bf k}_{\mu 1} &
c_\theta \,\rho^{\bf k}_{\mu 2} &- i s_\theta \,\lambda^{\bf k}_{\mu 1}
& i c_\theta \,\lambda^{\bf k}_{\mu 2}  \\ i c_\theta \,\lambda^{\bf
k}_{e1} & i s_\theta \,\lambda^{\bf k}_{e2} &c_\theta\, \rho^{\bf
k}_{e1} & s_\theta \,\rho^{\bf k}_{e2}  \\ - i s_\theta \,\lambda^{\bf
k}_{\mu 1} & i c_\theta\, \lambda^{\bf k}_{\mu 2} &- s_\theta\,
\rho^{\bf k}_{\mu 1} & c_\theta\, \rho^{\bf k}_{\mu 2} \\
\end{array}\right)
\left(\begin{array}{c} \alpha^{r}_{{\bf k},1}(t)\\ \alpha^{r}_{{\bf
k},2}(t)\\ \beta^{r\dag}_{{-\bf k},1}(t)\\ \beta^{r\dag}_{{-\bf k},2}(t)
\end{array}\right)\, ,
\end{equation}
where $c_\theta\equiv \cos\theta$, $s_\theta\equiv \sin\theta$ and
\begin{eqnarray}\label{rho}
 \rho^{\bf k}_{a b} \delta_{rs}&\equiv& \cos\frac{\chi_a - \chi_b}{2}
\delta_{rs}= u^{r\dag}_{{\bf k},a} u^{s}_{{\bf k},b} =
v^{r\dag}_{-{\bf k},a} v^{s}_{-{\bf k},b} \, ,
\\ \label{lambda} i
\lambda^{\bf k}_{a b}\delta_{rs} &\equiv& i \sin\frac{\chi_a -
\chi_b}{2} \delta_{rs} = u^{r\dag}_{{\bf k},a} v^{s}_{-{\bf k},b} =
v^{r\dag}_{-{\bf k},a} u^{s}_{{\bf k},b}\, ,
\end{eqnarray}
with $a,b = 1,2,e,\mu$. Since $\rho^{\bf k}_{1 2}=|U_{{\bf k}}|$ and
$i \lambda^{\bf k}_{1 2}=\epsilon^r |V_{{\bf k}}|$, etc.,
the operators (\ref{FHYmatrix})
reduce\footnote{Up to a minus sign, see Ref.~\cite{remarks}.} to the ones in Eqs.~(\ref{BVmatrix}) when $\mu_e\equiv m_1$ and
$\mu_\mu \equiv m_2$.

The generalized flavor vacuum, which is
annihilated by the flavor operators given by Eq.~(\ref{FHYmatrix}), is now
written as \cite{FHY99}:
\begin{equation}\label{FHYvac}
|{\widetilde 0(t)}\rangle_{e,\mu}\equiv
K^{-1}_{\theta,\mu}(t)|0\rangle_{1,2} ~.
\end{equation}
For $\mu_e\equiv m_1$ and $\mu_\mu \equiv m_2$, this state reduces to
the standard flavor vacuum $| 0(t)\rangle_{e,\mu}$  of Eq.~(\ref{2.22}).

\section{Gauge structure and choice of the representation\label{gauge}}

We have seen how it is possible to define exact
flavor charge eigenstates for mixed neutrinos.
We have also seen that an apparent arbitrariness exists in the
choice of the basis of free fields  respect to which the flavor fields are
expanded. Such a choice cannot be arbitrary since the structure of the flavor vacuum
and thus the physical results depend on it.
A remarkable result was presented in Ref.~\cite{remarks}, where it was shown that
the oscillation formulas are insensitive to the choice of such a basis.

However, there are other aspects which need to be considered. One is that Lorentz invariance
is broken, since
the flavor vacuum is explicitly time-dependent.
As a consequence, flavor states cannot be interpreted in terms
of irreducible representations of the Poincar\'{e} group.
A possible way to recover
Lorentz invariance for mixed fields  has been explored in
  Ref.~\cite{BlaMagueijo} where  non-standard dispersion relations
for the mixed particles have been related to  non-linear
realizations of the Poincar\'{e} group \cite{MagSmol}.

A different way has been explored in   Ref.~\cite{Blasone:2010zn}, where it has been shown that
a non-abelian gauge structure appears naturally in connection with flavor mixing.
In this framework, it is then possible to account for the above-mentioned
violation of Lorentz invariance due to the flavor vacuum  having, at the same time,
 standard dispersion relations for
flavor neutrino states.

To see how this is possible, let us note that  the Lagrangian Eq.~(\ref{lag2}) can be rewritten as
describing a doublet of Dirac fields in interaction with an
external Yang--Mills field:
\bea \label{LagrflavCov}{\cal L} = {\bar \nu}_f
(i\gamma^{\mu}D_{\mu} - M_d) \nu_f, \eea
where $\nu_f= (\nu_e, \nu_\mu )^T$ is the flavor doublet and
$M_d={\rm diag}(m_e,m_{\mu})$ is a diagonal mass matrix.
We have introduced a covariant derivative and a gauge field as
\bea
D_{\mu}&=& \partial_{\mu} + i\, g\, \beta\, A_{\mu},
\\ \label{Conn}
A_{\mu} &\equiv&\frac{1}{2} A_{\mu}^a  \si_a \,=\,  n_{\mu} \delta m \,\frac{\sigma_1}{2}\in
su(2), \qquad n^{\mu}\equiv(1,0,0,0)^T ,
\eea
where  $m_{e\mu}=\frac{1}{2}\,\tan 2\theta \,\delta m$, and
 $\delta m\equiv m_{\mu}-m_e$. We also define $g\equiv\tan 2\theta$  as the coupling constant for
the mixing interaction.
Flavor  mixing can thus be
seen as an interaction of the flavor fields
with an $SU(2)$ constant gauge field.

Here $\alpha_i$, $i=1,2,3$ and $\beta$ are the usual Dirac
matrices in a given representation. For definiteness,
we choose the following representation:
\bea \alpha_i=\lf(\ba{cc}0&\sigma_i\\\sigma_i&0\ea\ri),\qquad
\beta= \lf(\ba{cc}\ide&0\\0&-\ide\ea\ri),\eea
where $\sigma_i$ are the Pauli
matrices and $\ide$ is the $2\times 2$ identity matrix.

We now consider the energy momentum tensor
associated with the flavor neutrino fields
in interaction with the external gauge field \cite{Blasone:2010zn}:
\bea \label{newemtensor}\widetilde{T}_{\rho\sigma} &= &{\bar \nu}_f i
\gamma_{\rho}D_{\sigma}\nu_f - \eta _{\rho\sigma}{\bar \nu}_f
(i\gamma^{\lambda}D_{\lambda} - M_d)\nu_f\,,\eea
which  is to be compared with the canonical energy
momentum tensor associated with the Lagrangian
(\ref{lag2}):
\bea \label{emtensor}
 T_{\rho\sigma}&= &{\bar \nu}_f i
\gamma_{\rho}D_{\sigma}\nu_f - \eta _{\rho\sigma}{\bar \nu}_f
(i\gamma^{\lambda}D_{\lambda} - M_d)\nu_f
 + \eta_{\rho\sigma}m_{e\mu} {\bar \nu}_f \si_1\nu_f \,.
\eea

We then  define a $4$-momentum
operator as $\widetilde{P}^\mu \equiv \int d^3\mathbf{x} \,\widetilde{T}^{0\mu}$
and obtain a conserved $3-$momentum operator:
\bea \non
\widetilde{P}^i &=&  i \int d^3\mathbf{x}\,
\nu_f^{\dagger}\partial^i \nu_f
\\ \non
&=& i \int d^3\mathbf{x}\,
\nu_e^{\dagger}\partial^i \nu_e + i \int d^3\mathbf{x}\,
\nu_{\mu}^{\dagger}\partial^i \nu_{\mu}
\\  \label{Pgauge} &\equiv & \widetilde{P}^i_e(t) +
\widetilde{P}^i_{\mu}(t),
\qquad\qquad\qquad\qquad i=1,2,3 \eea
and a non-conserved Hamiltonian operator:
\bea \non \widetilde{P}^0(t)
&\equiv & \widetilde{H}(t) =
\int d^3 \mathbf{x} \,{\bar \nu}_f\lf( i  \gamma_0 D_0
- i \gamma^\mu D_\mu + M_d \ri)\nu_f
\\ \non
&=& \int d^3 \mathbf{x}\, \nu_e^{\dagger}\lf(-i\vect{\al}\cdot\vect{\nabla}+\beta
m_e\ri) \nu_e + \int d^3 \mathbf{x}
\,\nu_{\mu}^{\dagger}\lf(-i\vect{\al}\cdot\vect{\nabla}+\beta m_{\mu}\ri)
\nu_{\mu}
\\  \label{Hgauge}
&\equiv &\widetilde{H}_e(t) + \widetilde{H}_{\mu}(t).\eea
Note that both the Hamiltonian and the momentum  operators split
in a contribution involving only the electron
neutrino field and in another where only the muon neutrino field
appears.

We remark that the tilde Hamiltonian is \textit{not} the
generator of time translations.
This role competes to the complete
Hamiltonian $H=\int d^3\mathbf{x}\,T^{00}$,
obtained from the energy-momentum tensor
Eq.~(\ref{emtensor}).

We now show that it is possible to define
flavor neutrino states which are simultaneous eigenstates of the
$4$-momentum operators above constructed and of the flavor
charges. Such a non-trivial request requires
a redefinition of the flavor vacuum. Indeed,
one can show \cite{Blasone:2010zn} that this is achieved by means of the expansion
Eq.~(\ref{exnue1}), provided one sets $\mu_e=m_e$ and $\mu_\mu=m_\mu$.
From now on we use the tilde to denote the flavor operators so defined.

The tilde flavor operators are connected to those of Section~\ref{standardmix} by a
Bogoliubov transformation:
\bea \label{fujiiBog}
\lf( \ba{c} \widetilde{\alpha}_{\mathbf{k},\sigma}^r(t)\\
\widetilde{\beta}_{-\mathbf{k},\sigma}^{r\dagger}(t)\ea\ri)& =&
J^{-1}(t) \lf( \ba{c} \alpha_{\mathbf{k},\sigma}^r(t)\\
\beta_{-\mathbf{k},\sigma}^{r\dagger}(t)\ea\ri) J(t),
\\
J(t)&=&\prod_{\mathbf{k},r}\exp\lf\{
i\sum_{(\sigma,
j)}\xi^{\mathbf{k}}_{\sigma,j}\lf[\alpha_{\mathbf{k},\sigma}^{r \dagger }(t)
\beta_{-\mathbf{k},\sigma}^{r \dagger}(t) +
\beta_{-\mathbf{k},\sigma}^r(t)\alpha_{\mathbf{k},\sigma}^r(t)\ri]
\ri\},\label{NewGen}
\eea
with $ (\sigma,j)=(e,1),(\mu,2),$ and
with $\xi_{\sigma,j}^{\bf k}\equiv (\chi_\sigma - \chi_j)/2$ and
$\cot\chi_\sigma = |{\bf k}|/m_\sigma$, $\cot\chi_j = |{\bf k}|/m_j$.
The new (physical) flavor vacuum is given by
\bea\label{truevac} |{\widetilde 0}(t)\rangle_{e\mu}=J^{-1}(t)|
0(t)\rangle_{e\mu}.\eea

Notice that the
flavor charges are
invariant under the above Bogoliubov transformations \cite{remarks},
i.e.,  $\widetilde{Q}_{\sigma}=Q_{\sigma}$, with:
\bea
\widetilde{Q}_{\sigma}(t)=\sum_r \int
d^3\mathbf{k}\lf(\widetilde{\alpha}^{r \dagger}_{\mathbf{k}\sigma}(t)
\widetilde{\alpha}^r_{\mathbf{k}\sigma}
(t)- \widetilde{\beta}^{r \dagger}_{-\mathbf{k}\sigma}(t)
\widetilde{\beta}^r_{-\mathbf{k}\sigma}(t)\ri).\eea

In terms of the tilde flavor ladder operators, the Hamiltonian
and momentum operators Eqs.~(\ref{Pgauge}) and (\ref{Hgauge}) read:
\bea {}\hspace{-1cm}
\widetilde{\mathbf{P}}_\si(t)
&=& \sum_r\int d^3 {\bf k}\,\,  {\bf k} \left(
\widetilde{\alpha}^{r\dagger}_{{\bf k},\si}(t) \widetilde{\alpha}^{r}_{{\bf
k},\si}(t) + \widetilde{\beta}^{r\dagger}_{{\bf {k}},\si}(t)
\widetilde{\beta}^{r}_{{\bf k},\si}(t) \right),
\\ [2mm]
 {}\hspace{-1cm} \widetilde{H}_{\sigma}(t)&=&
\sum_r\int d^3{\bf k}\,  \om_{\mathbf{k},\si} \left(
\widetilde{\alpha}^{r\dagger}_{{\bf k},\si}(t)
\, \widetilde{\alpha}^{r}_{{\bf k},\si} (t)-
\widetilde{\beta}^{r}_{{\bf {k}},\si} (t)
\, \widetilde{\beta}^{r\dagger}_{{\bf k},\si} (t)\right)\!. \eea

Since all the above operators are diagonal, we can define
common eigenstates as follows:
\bea \label{NewFlavState}
|\widetilde{\nu}^{\, r}_{\mathbf{k},\sigma}(t)\rangle
= \widetilde{\alpha}^{r \dagger}_{\mathbf{k},\sigma}(t)
|\widetilde{0}(t)\rangle_{e\mu} \, ,\eea
and similar ones for the antiparticles.
We have
\bea
\lf(\ba{c}
\widetilde{H}_{\sigma}(t)\\\widetilde{\mathbf{P}}_{\sigma}(t)\ea\ri)
|\widetilde{\nu}^{\, r}_{\mathbf{k},\sigma}(t)\rangle
= \lf(\ba{c}
\omega_{\mathbf{k},\sigma}\\
\mathbf{k}\ea\ri)|\widetilde{\nu}^{\, r}_{\mathbf{k},\sigma}(t)\rangle,
\eea
making explicit the $4-$vector
structure.

Note that the above construction and the consequent Poincar\'e invariance
holds at a given time $t$. Thus, for each different time,
we have a different  Poincar\'e structure.
Flavor neutrino fields behave (locally in time) as ordinary on-shell fields with
definite masses $m_e$ and $m_{\mu}$, rather than
those of the  mass eigenstates of the standard approach,
$m_1$ and $m_2$. Flavor  oscillations then arise as a consequence of the
interaction with the (constant) gauge field, which acts as a birefringent medium and
can be seen as a \emph{neutrino aether}.

The operator $\widetilde{H}$  can be viewed as  the sum of the
kinetic energies of the flavor neutrinos, or equivalently as the
energy which can be extracted from flavor
neutrinos by scattering processes, the mixing energy being
``frozen''. A thermodynamical picture is given in Ref.~\cite{Blasone:2010zn}.

Finally we consider the condensation densities for the
physical flavor vacuum Eq.~(\ref{truevac}). They are given by
\bea \label{N1pfv}
\hspace{-.5cm} \,_{e,\mu}\lan\widetilde{0}(t)|\al_{{\bf k},1}^{r\dag}\al^r_{{\bf k},1}|\widetilde{0}(t)\ran_{e,\mu}&=&
\,_{e,\mu}\lan\widetilde{0}(t)|\bt_{{\bf k},1}^{r\dag}\bt^r_{{\bf k},1}|\widetilde{0}(t)\ran_{e,\mu}\,=\,
\cos^2\theta \sin^2\xi^{\bf k}_{e,1}\,+\,\sin^2\theta \sin^2\xi^{\bf k}_{e,2},
\\ \label{N2pfv}
\hspace{-.5cm} \,_{e,\mu}\lan\widetilde{0}(t)|\al_{{\bf k},2}^{r\dag}\al^r_{{\bf k},2}|\widetilde{0}(t)\ran_{e,\mu}&=&
\,_{e,\mu}\lan\widetilde{0}(t)|\bt_{{\bf k},2}^{r\dag}\bt^r_{{\bf k},2}|\widetilde{0}(t)\ran_{e,\mu}\,=\,
\cos^2\theta \sin^2\xi^{\bf k}_{\mu,2} + \sin^2\theta \sin^2\xi^{\bf k}_{\mu,1},
\eea
which have to be compared with the result Eq.(\ref{2.41b}).

\section{Phenomenology}

A number of (possible) physical effects  can be discussed starting from the results of last
Section. Here we briefly consider some of them.

\subsection{Neutrino oscillations}

The flavor oscillation formulas are derived
by computing, in the Heisenberg representation, the
expectation value of the flavor charge operators on the flavor state \cite{BHV99}.

In the physical representation above defined, we obtain\footnote{We use the notation
$|\widetilde{\nu}^{\, r}_{{\bf k},\rho}\rangle\equiv \widetilde{\alpha}^{r\dag}_{{\bf k},\rho}(0)|\widetilde{0}(0)\rangle_{e,\mu}$ and $|\widetilde{0}\rangle_{e,\mu}\equiv |\widetilde{0}(0)\rangle_{e,\mu}$. }
\bea\label{charge2}
\langle  \widetilde{\nu}^{\, r}_{{\bf k},\rho}|
\widetilde{Q}_\sigma(t) |\widetilde{\nu}^{\, r}_{{\bf k},\rho}\rangle
\,=\, \lf|\lf \{\widetilde{\alpha}^{r}_{{\bf k},\si}(t), \widetilde{\alpha}^{r
\dag}_{{\bf k},\rho}(0) \ri\}\ri|^{2} \;+ \;\lf|\lf\{\widetilde{\bt}_{{-\bf
k},\si}^{r \dag}(t), \widetilde{\alpha}^{r \dag}_{{\bf k},\rho}(0) \ri\}\ri|^{2},
\qquad \si,\rho=e,\mu.
\eea
A similar result holds if we consider expectation values on antiparticle states.

For the case of two flavors, the above formula turns out
to be identical \cite{remarks} to the one defined by
means of the representation of Section~\ref{standardmix}.
However, corrections arise for the case of three flavors, when CP violation is present \cite{3flav}.

\subsection{Beta decay}

In the above picture, flavor neutrinos states have standard dispersion relations and the oscillations are due to the interaction with the external gauge field (neutrino aether).
Consequently, in experiments for the direct measurement of the neutrino masses, as the ones based on tritium (beta) decay, flavor neutrinos are predicted to exhibit the behavior of ordinary free particles  with masses $m_e$ and $m_\mu$ rather than superpositions of massive states with masses
$m_1$ and $m_2$. This is discussed in detail in  Ref.~\cite{Blasone:2010zn}.

\subsection{Cosmological constant}

The energy of the flavor vacuum can provide a non-standard contribution to the
cosmological constant. In Ref.~\cite{Blasone:2004yh} such a contribution was evaluated
by means of the representation of Section~\ref{standardmix}.
In the light of the above results, it needs
to be recalculated by use of the correct flavor vacuum.

In Ref.~\cite{Blasone:2004yh}
the contribution $\lan\rho_{vac}^{mix}\ran$ of the neutrino
mixing to the vacuum energy density is shown to be:
 \bea\
 \lan\rho_{vac}^{mix}\ran \eta_{00}={}_{e,\mu}\lan \widetilde{0}  |\sum_{i} T^{(i)}_{00}|\widetilde{0} {\rangle}_{e,\mu}~,
 \eea
 where
\bea T^{(i)}_{00}= \ \sum_{r}\int d^{3}{\bf k}\,
\omega_{k,i}\lf(\al_{{\bf k},i}^{r\dag} \al_{{\bf k},i}^{r}+
\beta_{{\bf -k},i}^{r\dag}\beta_{{\bf -k},i}^{r}\ri), \quad i=1,2.\eea
By using the result Eqs.(\ref{N1pfv}), (\ref{N2pfv}), we obtain
\bea\label{34} \lan\rho_{vac}^{mix}\ran \eta_{00}
&=& 4 \int d^{3}{\bf k}\, \omega_{k,1} \lf(\cos^2\theta \sin^2\xi^{\bf k}_{e,1}\,+\,\sin^2\theta \sin^2\xi^{\bf k}_{e,2}\ri)\\
&+& 4 \int d^{3}{\bf k}\, \omega_{k,2}\lf( \cos^2\theta \sin^2\xi^{\bf k}_{\mu,2} + \sin^2\theta \sin^2\xi^{\bf k}_{\mu,1}\ri). \eea
which is to be compared with the result of Ref.~\cite{Blasone:2004yh}. Notice that the above
contribution is zero in the no-mixing limit when the mixing
angle $\theta = 0$ and/or $m_{1} = m_{2}$.

When a cutoff $K$ is introduced, the above integral can be evaluated and the result
 turns out to be
proportional to $K^2$,
whereas the usual free-field zero-point energy contribution would be going
like $K^4$.

\subsection{Supersymmetry}

It has been argued  \cite{Capolupo:2010ek,Mavromatos:2010ni} that the flavor vacuum structure can provide a mechanism for
supersymmetry breaking. This can be seen \cite{Capolupo:2010ek} by considering a Lagrangian of  the form\footnote{The symbol $M_d$ introduced here should not be confused with the one of
Section \ref{gauge}.}
\bea\label{MixLagrangian}\non
\mathcal{L}&= &-\frac{i}{2}\,\bar{\psi}_f\,( \not\!\partial + M)\,\psi_f \,-\, \frac{1}{2}\,
\partial_{\mu}S_f \,\partial^{\mu}S_f \,
-\,\frac{1}{2} \,S_f^{T}\, M^2 \,S_f \,-\, \frac{1}{2}\,\partial_{\mu}P_f \,\partial^{\mu}P_f \,-\,
\frac{1}{2} \,P_f^{T} \,M^2 \,P_f,
\\
 &=& - \frac{i}{2}\,\bar{\psi}\,( \not\!\partial + M_d)\,\psi \,-\,
 \frac{1}{2}\,\partial_{\mu}S \,\partial^{\mu}S \,-\,
 \frac{1}{2} \,S^{T}\, M_d^2 \,S \,-\,
  \frac{1}{2}\,\partial_{\mu}P \,\partial^{\mu}P \,-\,\frac{1}{2}\, P^{T} \,M_d^2 \,P\,,
\eea
with $M=\lf(\begin{array}{cc} m_a&m_{ab}\\ m_{ab} &m_b\end{array}\ri)$ and $M_d=\textrm{diag}(m_1,m_2)$. The fields are two free Majorana fermions $\psi_i$, two free real scalars $S_i$, two free real pseudoscalars $P_i$:  $\psi=(\psi_1,\psi_2)^T, S=(S_1,S_2)^T, P=(P_1,P_2)^T$.
The flavor fields are defined as:
\bea\label{mixing-transf}
\psi_f= U \psi, \qquad S_f = U S,\qquad P_f= U P,
\eea
where $\psi_f=(\psi_a, \psi_b)^T$ ,  $S = (S_a,S_b)^T$, $P = (P_a,P_b)^T$, and $U= \lf(\begin{array}{cc} \cos\theta&\sin\theta\\ -\sin\theta &\cos \theta\end{array}\ri)$.
 It is also $m_a = m_1\cos^2\theta + m_2\sin^2\theta$, $m_b = m_1\sin^2\theta + m_2\cos^2\theta$, and $m_{ab}=(m_2-m_1)\sin\theta\cos\theta$.

The Fourier expansion of the fields ($i=1,2$) are:
\bea
\psi_i(x)&=& \sum_{r=1}^2\int \frac {d^3\mathbf{k}}{(2\pi)^{\frac{3}{2}}}\,\, e^{i \mathbf{k}\mathbf{x}}\lf[u^r_{\mathbf{k},i}\alpha^r_{\mathbf{k},i}e^{-i\omega_{k,i}t}
+ v^r_{-\mathbf{k},i}\alpha^{\dagger
r}_{\mathbf{-k},i}e^{i\omega_{k,i}t}\ri],\\
S_i(x)&=& \int \frac {d^3\mathbf{k}}{(2\pi)^{\frac{3}{2}}}\,\,
\frac{1}{\sqrt{2 \omega_{k,i}}}  e^{i \mathbf{k}\mathbf{x}}
\lf[b_{\mathbf{k},i} e^{-i\omega_{k,i}t}
+ b^{\dagger}_{\mathbf{-k},i}e^{i\omega_{k,i}t}\ri],\\
P_i(x)&=& \int \frac {d^3\mathbf{k}}{(2\pi)^{\frac{3}{2}}}\,\,
\frac{1}{\sqrt{2 \omega_{k,i}}}  e^{i \mathbf{k}\mathbf{x}}
\lf[c_{\mathbf{k},i} e^{-i\omega_{k,i}t}
+ c^{\dagger}_{\mathbf{-k},i}e^{i\omega_{k,i}t}\ri],
\eea
where $v^r_{\mathbf{k},i}=\gamma_0 C (u^r_{\mathbf{k},i})^*$ and $u^r_{\mathbf{k},i}=\gamma_0 C (v^r_{\mathbf{k},i})^*$ by the Majorana condition  and the operators $\alpha^r_{\mathbf{k},i}$, $b_{\mathbf{k},i}$ and $c_{\mathbf{k},i}$ annihilate the vacuum $|0\rangle=|0\rangle^{\psi}\otimes|0\rangle^S\otimes|0\rangle^P$. The expectation value of the Hamiltonian on this vacuum is zero
\bea
\langle 0|(H_{\psi}+ H_B)|0\rangle=0,
\eea
where $H_B=H_S+H_P$.
The  generators of the mixing transformations (\ref{mixing-transf})
 are given by \cite{Blasone:2003hh}:
\bea
G_{\psi}(\theta) &=& \exp\lf[\frac{\theta}{2}\int d^3 \mathbf{x}\lf(\psi_1^{\dagger}(x)\psi_2(x) - \psi_2^{\dagger}(x)\psi_1(x)\ri) \ri],\\
G_S(\theta) &=& \exp\lf[-i\theta \int d^3 \mathbf{x}(\pi_1^S(x) S_2(x)- \pi_2^S(x) S_1(x)) \ri],\\
G_P(\theta) &=& \exp\lf[-i\theta \int d^3 \mathbf{x}(\pi_1^P(x) P_2(x)- \pi_2^P(x) P_1(x)) \ri],
\eea
where  $\pi_i^S(x)$ and $\pi_i^P(x)$ are the conjugate momenta of the fields $S_i(x)$ and $P_i(x)$, respectively.
The flavor vacuum (at $t=0$)  is:
$|0\ran_{f}\,\equiv\,|0\ran_{f}^{\psi}\, \otimes \,|0\ran_{f}^S\,
\otimes \,|0\ran_{f}^{P}\, $, where
\bea
\label{mixed-vacua}
|0\ran_{f}^{\psi}\,\equiv \, G^{-1}_{\psi }(\te) \; |0\ran^{\psi}\,,
\qquad
|0\ran_{f}^S\,\equiv \, G^{-1}_S(\te) \; |0\ran^S\,,
\qquad
|0\ran_{f}^{P}\,\equiv \, G^{-1}_{P}(\te) \; |0\ran^{P}\,,
\eea
are the flavor vacua of the fields $\psi_{\si}(x)$, $S_{\si}(x)$, $P_{\si}(x)$,
respectively.

In a similar way as done above, we now introduce the physical flavor vacuum:
\bea |\widetilde{0}\ran_{f}\,\equiv\,|\widetilde{0}\ran_{f}^{\psi}\, \otimes \,|\widetilde{0}\ran_{f}^S\,
\otimes \,|\widetilde{0}\ran_{f}^{P}\,,\eea
The expectation value of the fermionic part of $H$ on  $|\widetilde{0}\ran_{f}$
is given by:
\bea\label{Hflav}\non
{}_{f}\lan \widetilde{0}| H_{\psi} |\widetilde{0} \ran_{f}\,
&=&-\,  \int d^{3}{\bf k} \,
(\omega_{k,1} + \omega_{k,2}) \,
\\ \non
&+&2 \int d^{3}{\bf k}\, \omega_{k,1} \lf(\cos^2\theta \sin^2\xi^{\bf k}_{a,1}\,+\,\sin^2\theta \sin^2\xi^{\bf k}_{a,2}\ri)\\
&+& 2 \int d^{3}{\bf k}\, \omega_{k,2}\lf( \cos^2\theta \sin^2\xi^{\bf k}_{b,2} + \sin^2\theta \sin^2\xi^{\bf k}_{b,1}\ri),  \eea
while for the bosonic part we obtain:
\bea\label{HflavBos} \non
{}_{f}\lan \widetilde{0}| H_B | \widetilde{0} \ran_{f}
&=&  \int d^{3}{\bf k} \,
(\omega_{k,1} + \omega_{k,2}) \,
\\ \non
&+&2 \int d^{3}{\bf k}\, \omega_{k,1} \lf(\cos^2\theta \sin^2{\hat \xi}^{\bf k}_{a,1}\,+\,\sin^2\theta \sin^2{\hat \xi}^{\bf k}_{a,2}\ri)\\
&+& 2 \int d^{3}{\bf k}\, \omega_{k,2}\lf( \cos^2\theta \sin^2{\hat \xi}^{\bf k}_{b,2} + \sin^2\theta \sin^2{\hat \xi}^{\bf k}_{b,1}\ri),   \eea
with ${\hat \xi}_{\sigma,i}^{\,\bf k}\equiv \frac{1}{2}
\ln\frac{\om_{k,\si}}{\om_{k,i}}$ (see Ref.~\cite{Blasone:2001du}).

Combining Eqs.(\ref{Hflav}) and (\ref{HflavBos}) we finally have:
\bea\label{violation}\non
&& \hspace{-1cm}\, {}_{f}\lan \widetilde{0}| (H_{\psi}+ H_B) | \widetilde{0} \ran_f\,=\,
2 \ \cos^{2}\theta \, \int d^{3}{\bf k} \, \omega_{k,1} \lf(\sin^2\xi^{\bf k}_{a,1} +\sin^2{\hat \xi}^{\bf k}_{a,1} \ri) + \omega_{k,2} \lf(\sin^2\xi^{\bf k}_{b,2} + \sin^2{\hat \xi}^{\bf k}_{b,2}\ri)
\\
&&\hspace{2cm}+ \
2 \ \sin^{2}\theta \, \int d^{3}{\bf k} \, \omega_{k,1} \lf(\sin^2\xi^{\bf k}_{a,2} +\sin^2{\hat \xi}^{\bf k}_{a,2} \ri) + \omega_{k,2} \lf(\sin^2\xi^{\bf k}_{b,1} + \sin^2{\hat \xi}^{\bf k}_{b,1}\ri)
\,,
\eea
which exhibits
supersymmetry breaking  associated to flavor mixing. The above result differs from that of Ref.~\cite{Capolupo:2010ek}.

\section{Conclusions}

In the framework of the quantum field theory treatment of particle mixing, we have discussed flavor states for two flavor neutrino mixing. We have given arguments for selecting a physically relevant representation for the flavor states. Phenomenological consequences of our discussion have been explored, also in connection to previous results.

\ack We  acknowledge partial financial support from MIUR and INFN.

\end{document}